# ME-Net: Multi-Encoder Net Framework for Brain Tumor Segmentation

Wenbo Zhang [1,*], Guang Yang [2, 3, *†], He Huang [1], Weiji Yang[4], Xiaomei Xu[1], Yongkai Liu [5], Xiaobo Lai [1,†]

[1] College of Medical Technology, Zhejiang Chinese Medical University, Hangzhou, 310053, China
[2] Cardiovascular Research Centre, Royal Brompton Hospital, London, SW3 6NP, UK
[3] National Heart and Lung Institute, Imperial College London, London, SW7 2AZ, UK
[4] College of Life Science, Zhejiang Chinese Medical University, Hangzhou, 310053, China
[5] Department of Radiological Sciences, David Geffen School of Medicine, University of California at Los Angeles, Los Angeles, CA, USA

[*] Co-first authors contributed equally.
[†] Corresponding Authors: G. Yang (g.yang@imperial.ac.uk) and X. Lai (dmia_lab@zcmu.edu.cn)

**Abstract**

Glioma is the most common and aggressive brain tumor. Magnetic resonance imaging (MRI) plays a vital role to evaluate tumors for the arrangement of tumor surgery and the treatment of subsequent procedures. However, the manual segmentation of the MRI image is strenuous, which limits its clinical application. With the development of deep learning, a large number of automatic segmentation methods have been developed, but most of them stay in 2D images, which leads to subpar performance. Moreover, the serious voxel imbalance between the brain tumor and the background as well as the different sizes and locations of the brain tumor makes the segmentation of 3D images a challenging problem. Aiming at segmenting 3D MRI, we propose a model for brain tumor segmentation with multiple encoders. The structure contains four encoders and one decoder. The four encoders correspond to the four modalities of the MRI image, perform one-to-one feature extraction, and then merge the feature maps of the four modalities into the decoder. This method reduces the difficulty of feature extraction and greatly improves model performance. We also introduced a new loss function named "Categorical Dice", and set different weights for different segmented regions at the same time, which solved the problem of voxel imbalance. We evaluated our approach using the online BraTS 2020 Challenge verification. Our proposed method can achieve promising results in the validation set compared to the state-of-the-art approaches with Dice scores of 0.70249, 0.88267, and 0.73864 for the intact tumor, tumor core, and enhanced tumor, respectively.

# 1. Introduction

Glioma is one of the most common primary brain tumors, accounting for 70% of adult malignant primary brain tumors. According to the malignant degree of glioma, it can be divided into low-grade glioma (LGG) and high-grade glioma (HGG) (S. Bakas et al., 2018). This kind of tumor is invasive, mainly spreading along with the white matter fiber bundles, damaging the healthy tissues of the brain (Hussain et al., 2017), and the growth rate is fast, difficult to find. The usual treatment methods include surgery, radiation therapy, etc. However, the incidence of glioma is high, and the therapeutic effect is poor. Once diagnosed, the survival time of the patient does not exceed 14 months (Pereira et al., 2016), therefore early diagnosis and treatment are crucial.

At present, Magnetic Resonance Imaging (MRI) has been widely used by radiologists as an important tool for the early diagnosis and treatment of brain tumors (Myronenko et al., 2019). MRI can visualize the site of interest and can generate multi-modal images to provide more tumor information. For example, 3D MRI can obtain sequence information of brain tumors in different spatial locations through one scan, including four sequences, i.e., fluid attenuation inversion recovery (FLAIR), T1 weighted (T1), T1 weighted contrast enhancement (T1-CE), and T2 weighted (T2). However, manual segmentation of brain tumors from MRI images requires considerable time, resources, and expertise (Zhou et al., 2019). Moreover, because manual segmentation relies on clinical experience, different experts have significant differences in segmenting each tumor subregion (B. H. Menze et al., 2015). Automatic segmentation by computer not only saves time and cost but also improves the objectivity of quantitative analysis (Stefano et al., 2020). Therefore, an automatic brain tumor segmentation tool is urgently needed to solve this problem. In recent years, with the development of artificial intelligence, deep learning has been widely used in the field of medical image segmentation and has become the mainstream method with its advantages of high precision and efficiency.

So far, many outstanding studies have been proposed for general medical segmentation algorithms, such as the U-Net structure proposed by Ronneberger et al (Ronneberger et al., 2015). and the V-Net structure proposed by Milletari et al (Milletari et al., 2016). These methods are end-to-end networks, including encoders and decoders. Specifically, the encoder is used to extract the features of the input picture, and the decoder is used to decompress the picture information. For multi-modal brain tumor segmentation, traditional V-Net can mix the four modalities of brain tumors for processing, but different modalities share the same weight and process the same. This method is not conducive to extracting the different features contained in each modal, so we believe that V-Net is not suitable for multi-modal image segmentation tasks. Based on this, we propose a novel network structure with four encoders called Multi-Encoder Net (ME-Net) for the four modalities of MRI images. Our method does not learn the characteristics of each mode, and fuses the four modes in the later stage. This is achieved by using the number of encoders corresponding to the number of modes in the encoder stage. This architecture can process multiple input images simultaneously and extract specific features. Its network structure is shown in Fig. 1. The contributions of this study are: 1) Propose a multi-encoder segmentation model. 2) Solve the problem of multi-modal input for brain tumors. 3) A better segmentation result is obtained.

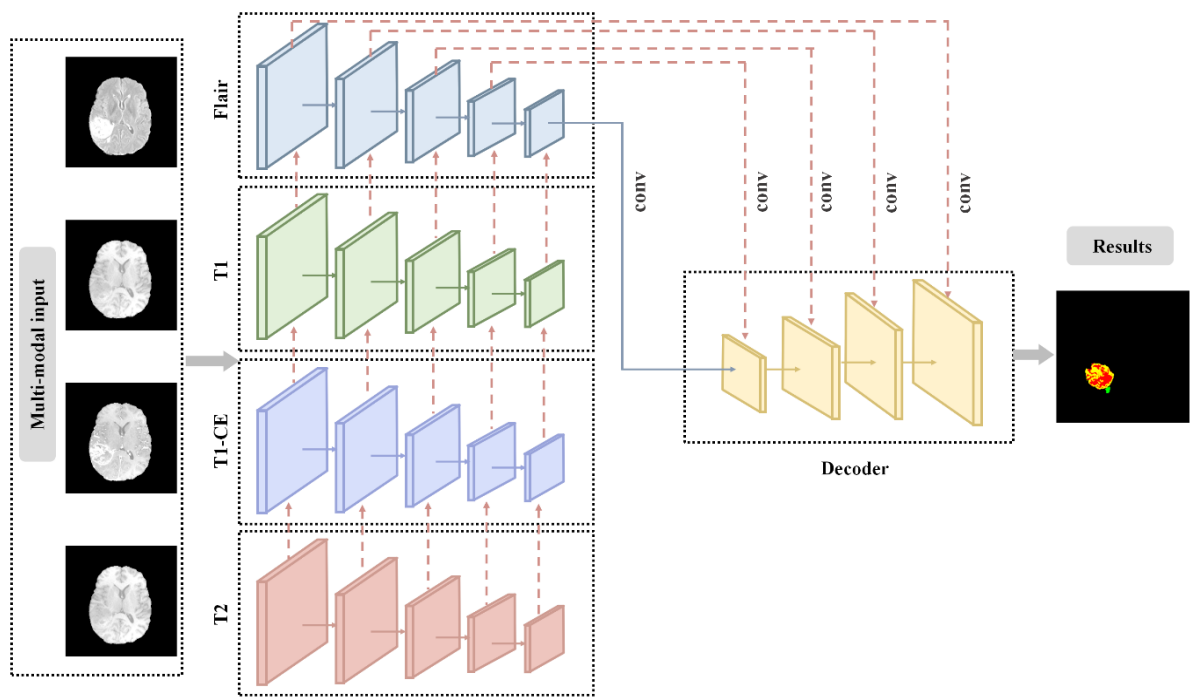

**Fig. 1.** Schematic diagram of our proposed ME-Net.

## 2. Related Works

*2.1. Traditional machine learning*

In the early stage of image segmentation, deep learning has not yet emerged, and traditional machine learning methods are often used for segmentation. The earliest machine learning method used for segmentation is the algorithm based on a threshold. Joseph et al. used thresholds to determine lung tissue, thus realizing automatic segmentation applied to X-ray computed tomography [(Joseph K et al., 2003)](). The threshold-based segmentation method directly used the grayscale characteristics of the image, so the calculation speed could be fast, but it could only be suitable for images with a large difference between the target and the background. There was a more commonly used segmentation method that combines edge segmentation and region segmentation. For the problem of brain protein structure detection, Tang et al. first adopted edge segmentation, and then adopted region segmentation based on the connectivity of the brain tissue structure to show more refined brain tissue structure [(Tang et al., 2000)](). However, this method also had limitations and could only be used for images with regional structures. The k-means algorithm had been widely used since its occurrence. Sulaiman et al. applied the k-means clustering algorithm to MR breast image segmentation [(Sulaiman S.N et al., 2010)](). Juang et al. proposed a k-means-based color conversion segmentation algorithm [(Juang et al., 2010)](). The core idea was to input the grayscale image converted into a color space image, and then the image marked by the clustering index was operated to segment the given MRI image. The disadvantage of this method was that it did not consider spatial information and was sensitive to noise and gray-scale unevenness.

*2.2. Deep learning*

In recent years, with the rapid development of artificial intelligence, deep learning has begun to be widely used in the field of medical image segmentation. Wu et al. proposed a cell membrane segmentation method based on an iterative convolutional neural network, which improved the accuracy of cell membrane segmentation [(Wu et al., 2015)](). Long et al. proposed a fully convolutional neural network (FCN) based on a convolutional neural network (CNN) to segment medical images [(Long et al., 2014)](). They replaced the fully connected layer (FCN) with the upper convolutional layer. FCN could classify images at the pixel level, thereby solved the problem of semantic image segmentation. Ronneberger et al. proposed a fully convolutional network called U-Net for medical image segmentation (Ronneberger et al., 2015). Different from the FCN network, the up-sampling stage and the down-sampling stage of U-Net used the same number of convolutions, and the skip connection structure was used to connect the down-sampling layer and the up-sampling layer, which made the U-Net network segmentation accuracy higher. Milletari et al. proposed a fully convolutional neural network for volumetric medical image segmentation called V-Net, unlike U-Net, a three-dimensional convolution kernel was used in V-Net, and an objective function based on maximizing Dice coefficient was proposed to optimize the model (Milletari et al., 2016). Recurrent neural networks (RNN) had achieved great success in many natural language processing problems in recent years (Graves A et al., 2003). Stollenga et al. used 3D LSTM-RNN to segment brain MR images in six directions for the first time, rearranging the traditional cuboid calculation order in MD-LSTM using a pyramid method, and achieved good results in the MRBrainS challenge (Stollenga M. F. et al., 2015). Comelli A. et al. proposed the use of ENet and ERFNet for automatic segmentation of ascending thoracic aortic aneurysms. These two models were originally used for image segmentation of autonomous vehicles. They also show good performance in medical images. With the same accuracy, ENet is much faster than U-Net (Comelli A. et al., 2020). At the same time, Comelli A. et al. also used E-Net to segment CT images of idiopathic pulmonary fibrosis. Compared with U-Net, E-Net showed better performance (Comelli A. and Claudia C. et al., 2020).

*2.3. Brain tumor segmentation*

There are a large number of brain tumor segmentation algorithms, and most of them are based on the improvement of the basic deep learning framework. The accuracy is important not only for diagnosis but also for subsequent treatment. Hussain et al. proposed an automatic brain tumor segmentation algorithm based on Deep Neural Convolutional Network (DCNN), using a patch-based method, combined with the inception module, to extract two different sizes from the input image concentric patch to train network depth (Hussain S. and Anwar S.M. et al., 2017). Through integrated modeling, Xue et al. trained multiple 3D U-Net networks with different numbers of encoder blocks and decoder blocks, different input slice sizes, and different loss weights to segment brain tumors to achieve better segmentation results (Xue et al., 2018). Dmitry et al. proposed a deep cascade method for the automatic segmentation of brain tumors. The U-Net architecture was modified which can effectively process multi-modal MRI input. Besides, they also introduced a way to improve the quality of segmentation by using context obtained from the same topology model operated on scaled-down data (Dmitry et al., 2018). McKinly et al. embed the DenseNet structure with dilated convolution into a network similar to U-Net, introducing a new loss function, binary entropy generalization, to resolve label uncertainties. This approach reduced overfitting in segmentation (McKinley et al., 2018). Zhou et al. designed a variety of deep architectures with different structures to learn context and information of interest, and then integrated the predictions of these models to obtain more robust segmentation results (Zhou and Chen et al., 2019). Wang et al. proposed an end-to-end automatic segmentation method called a wide residual and pyramid pool network (WRN-PPNet). First, a two-dimensional slice was obtained from a three-dimensional MRI brain tumor image and input into the WRN-PPNet model to achieve the segmentation result (Wang and Li et al., 2019).

2.4. Our work

Although various deep learning frameworks have been proposed and significant progress has been made in the task of brain tumor segmentation, there are still challenges in this field. Firstly, the location, morphology, and size of gliomas vary significantly between patients, increasing the difficulty of segmentation (Prastawa et al., 2004). Second, because the lesion area scanned by MRI is very small in many cases, it causes a class imbalance between the lesion area and the background area (Chen et al., 2018). Finally, there is also an imbalance in the intensity of MRI images, such as the inconsistent sequence and intensity range of the scanner (L.G. Nyul et al., 2000). These problems affect the accuracy of the segmentation of brain tumors.

In this paper, in order to meet the above challenges, we propose a novel network structure called Multi-Encoder Network (ME-Net), which is an improvement on the V-Net structure proposed by Milletari et al. The traditional V-Net has only one down-sampling path, and it is difficult to extract MRI images of four modalities. To solve this problem, we have designed a new architecture composed of four encoders with similar structures. Its network structure is shown in Fig. 1. The input of the four down-sampling paths corresponds to the four modalities of brain tumors. This one-to-one idea greatly improves the feature extraction ability of the model. The corresponding modules of the four encoders are spliced by skip connection and finally merged into a feature map and input to the decoder. In addition, in the up-sampling process, the feature maps of the corresponding down-sampling stage are also integrated to make up for the lost information. We also design a Categorical Dice loss function as the optimization function of the model and adjust the weight value of each output category through the weight distribution, thereby solving the problem of the imbalance of foreground and background voxels.

We tested the performance of our proposed model on the Multi-modal Brain Tumor Segmentation Challenge (BraTS) 2020 dataset and compared it with the results of other teams participating in the challenge. The results show that our model has a promising performance for brain tumor segmentation. The innovations of the method in this article are: 1) This paper proposes a novel network structure with multiple encoders. 2) One-to-one feature extraction is performed on MRI images of four modalities, which greatly improves the segmentation performance. 3) Development of the Categorical Dice loss function to solve the problem of foreground and background voxel imbalance.

## 3. Methods

Our task is to segment 3D MRI brain tumor images containing four modalities. Multi-category segmentation has always been a huge challenge in the field of computer vision. To obtain good segmentation performance, a powerful network structure is often required. For this reason, we propose a novel network structure named ME-Net, which is mainly composed of down-sampling and up-sampling. We show the network structure of ME-Net in Fig. 1. In this section, we will introduce in detail the process of preprocessing, the network structure of ME-Net and the loss function used.

*3.1. Data preprocessing and data augmentation*

Since MRI has four modalities, FLAIR, T1, T1-CE, and T2, the image contrast of these modalities are all different, which may cause the phenomenon of gradient disappearance during training. Standardization can reduce the generation of gradient disappearance, therefore, we use the z-score method to standardize the images of each mode separately, and process the data by subtracting the average value from the image pixels and dividing by the standard deviation. Fig. 2 shows the comparison of the four modes of a sample before and after preprocessing. Standardization makes the characteristics of the tumor more obvious, which is helpful to improve the accuracy of segmentation. The standardized formula is as follows:

$$\hat{X} = \frac{X - \mu}{\sigma},\qquad(1)$$

where $X$ donates the image matrix, $\mu$ donates the mean of the image, $\sigma$ donates standard deviation, and $\hat{X}$ donates the normalized image matrix.

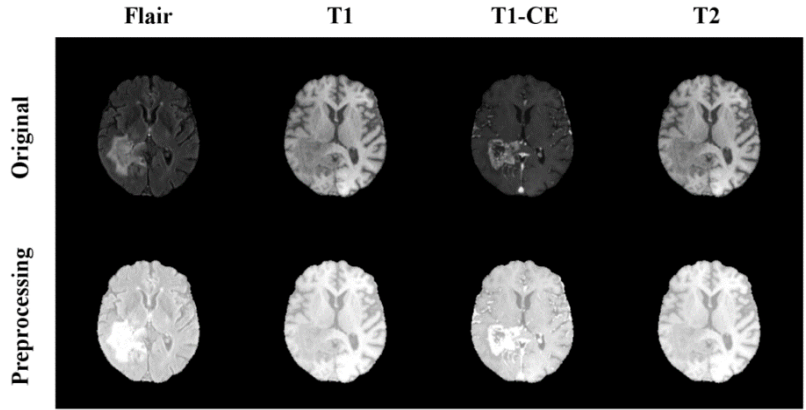

**Fig. 2.** Comparison before and after image standardization.

After standardization, for each case, four modal images with the same contrast are fused to form a three-dimensional image with four channels. The original image size is 240×240×155×1, and the combined image size is 240×240×155×4. Then perform a one-hot encoding operation on the mask image to generate the original image size 240×240×155×1 into 240×240×155×4, where the non-zero value area in channel 0 is the background area, the non-zero value in channel 1 is the necrosis area, and the non-zero value in channel 2 is the edema Area, the non-zero value in channel 3 is to enhance the tumor area.

Data augmentation can improve network performance and reduce the occurrence of overfitting. The data augmentation method we use is the patch operation, which only processes one patch at a time, rather than the entire image, in order to better detect edge features and is also a good regularization method. We divide the image and the mask into multiple blocks, a case generates 175 images with a size of 128×128×64. Then re-split the 4 channels of the image as the input of the four encoders of the model.

*3.2 ME-Net framework*

*3.2.1. Downsampling*

In ME-Net, we designed a special down-sampling structure for brain tumor segmentation tasks. The entire down-sampling path is composed of four encoders with similar structures. The inputs of these four encoders correspond to the four modalities FLAIR, T1, T1-CE, and T2 of the brain tumor. Use skip-connection to combine the modal features extracted by each encoder. Compared with the traditional V-Net using only one encoder to extract features from four modalities, the difficulty of extracting image features by the encoder can be reduced by using a specific encoder to extract features from a specific modal. This one-to-one feature extraction idea greatly improves the feature extraction ability of the model downsampling process.

Encoder adopts the structure and idea similar to V-Net, which function is to compress the image. The structure diagram of the encoder block is shown in Fig. 3. Each encoder block consists of 1 to 3 convolutional layers and a downsampling layer. The convolution formula is shown in (2) and (3),

$$i_s = i + (s-1)(i-1) ,  \quad (2)$$

$$o = \left[ \frac{i_s + 2p - k}{s} + 1 \right] ,  \quad (3)$$

where $i$ represents the input size, $i_s$ represents the input size after padding, $k$ represents the size of the convolution kernel, $p$ represents boundary expansion and $o$ represents the output size. $[\cdot]$ is the largest integer that does not exceed this value.

Here we use a convolutional layer with a step size of 2 to replace the pooling layer. This operation can double the number of channels and halve the resolution while maintaining computational complexity. The size of the convolution kernel in the encoder is 3×3×3 and the convolutional layer applies Batch normalization and ReLU activation function, except that the last layer of the network uses the sigmoid activation function, the rest is the same. In addition, the idea of ResNet short-circuiting is quoted inside each encoder block to avoid the gradient disappearing due to the excessively deep network structure. The characteristic diagrams of the four modalities of brain tumors are shown in Fig. 4 (a).

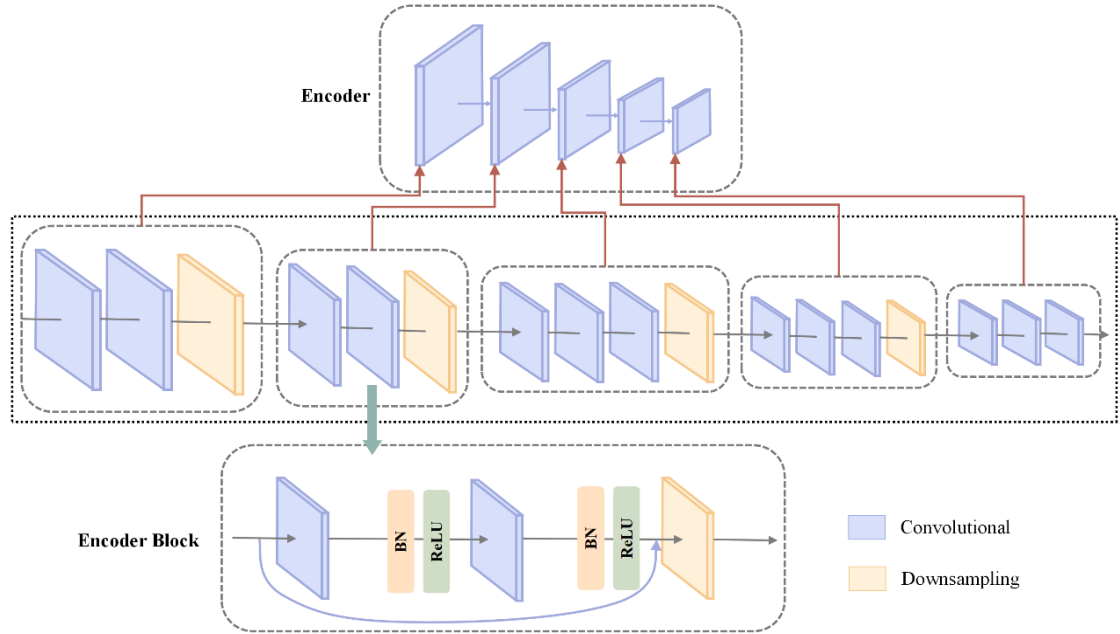

**Fig. 3.** An encoder branch structure diagram.

*3.2.2 Upsampling*

The structure of the decoder is similar to that of the encoder, each upsampling stage consists of 2 to 3 layers of convolutional layers and an upsampling layer. The function of the decoder is to decompress the image information. In upsampling, we use deconvolution with a step size of 2 to expand the resolution of the input image. The formula for deconvolution is shown in (4). After passing through a decoder, the size of the feature map is doubled, and the channel number of feature maps is reduced by half. The feature map of upsampling is shown in Fig. 4 (b).

$$o = s(i-1) + 2p - k + 2. \tag{4}$$

Each stage receives the characteristics of the down-sampling corresponding stage. The last convolutional layer of downsampling uses a convolution kernel with a size of 1×1×1 to keep the number of output channels consistent with the number of categories. Finally, the Softmax layer is used to convert the value of each channel into a probability value to realize the idea of the pixel by pixel segmentation. The idea of ResNet short-circuiting is also adopted in each decoder block.

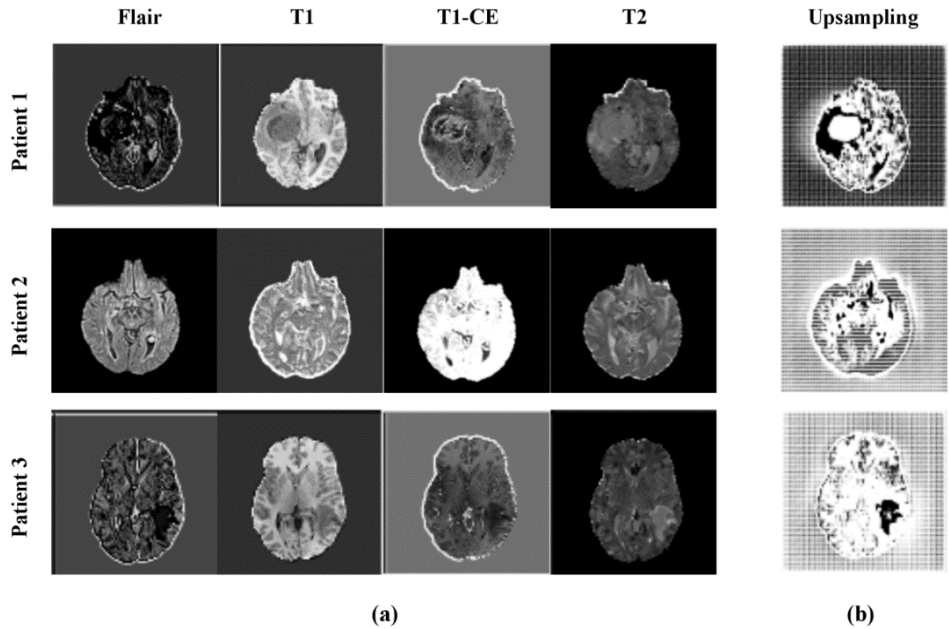

**Fig. 4.** Partial feature map display. Each row in the figure represents a sample, (a) is the output of the four modes after the first layer of downsampling, (b) is the output of the four modes after the first layer of upsampling is fused.

*3.2.3. Fusion strategy*

The fusion of the feature maps in ME-Net mainly includes two aspects, one is the fusion of the feature maps of the corresponding stages of the four encoders, and the other is the fusion of the feature maps of the corresponding stages of the encoder and the decoder. In the downsampling process, we use skip-connection to combine the modal features extracted by each encoder. As shown in Fig. 1, each encoder branch is similar to the encoder in V-Net, and they will be trained to get different weights. In each stage of downsampling, the feature map output from this stage is passed to the same stage of the parallel encoder using skip connections, and the feature map is also used as the input of the next stage of downsampling. After continuous iterative fusion in the down-sampling stage, we finally obtain feature maps with four modal fusions in each stage. These feature maps effectively combine the feature information of the four modalities in each stage of down-sampling. Then, perform 1×1×1 convolution operation on the feature maps obtained by combining different stages, change the number of channels, and pass them to the corresponding stage of upsampling. In the entire network, we use a large number of skip-connection operations to collect the fine-grained details that may be lost in the four compression paths. This method not only improves the quality of the final contour prediction but also proves to speed up the convergence time of the model.

In order to further make up for the information lost in the downsampling of the encoder, the skip-connection is used between the encoder and the decoder of the network to fuse the feature maps at the corresponding positions in the two processes, so that the decoder can obtain the information during upsampling. With more high-resolution information, the detailed information in the original image can be restored more perfectly. This cross-layer feature reconstruction module is based on the encoder-decoder structure. As the network continues to deepen, the detailed information retained by the feature map becomes less and less. Based on the encoder-decoder corresponding structure, skip-connection is used to splice the feature maps extracted from down-sampling during the encoder process and the new features obtained from up-sampling during the decoder process to preserve more important feature information to achieve a better segmentation effect.

### 3.3. Loss function

Medical image segmentation often faces the problem of extremely uneven foreground and background regions, and our brain tumor segmentation task also faces this challenge. For this reason, we improved the Generalized Dice Loss function (GDL) proposed by Sudre C.H. et al. and proposed a Categorical Dice loss function. Generalized Dice has been proven to be a good loss function in solving the imbalance problem of brain tumors [(Sudre C.H. et al., 2017)](#), and its calculation formula is shown in (5),

$$\text{GDL} = 1 - 2\frac{\sum_{l=1}^{2}\omega_{l}\sum_{i=1}^{N}p_{ln}g_{ln}}{\sum_{l=1}^{2}\omega_{l}\sum_{i=1}^{N}p_{ln}+g_{ln}}, \tag{6}$$

The weight is defined as the reciprocal of the square of the volume of each label, that is $\omega_{l} = 1/(\sum_{n=1}^{N}g_{ln})^{2}$. Where N represents all voxels, $l$ represents the number of categories, $p$ represents the predicted voxel, $g$ represents the real voxel.

We believe that because the background area of the data set we use accounts for more than 99% and this way of weight definition tends to cause the weight of the background class to tend to a minimum, we choose to give each class a fixed weight. We set the weight of the background to 0.1, and set the weight of necrosis, edema, and enhanced tumor to 1, and adjust the front and background imbalance through this weight distribution method. The Dice coefficient is used to calculate the similarity between the mask and the prediction, the range is [0, 1], and the formula is shown in (6),

$$\text{Dice}(P,G) = \frac{2|P \cap G|}{|P|+|G|} = \frac{2\sum_{i=1}^{N}p_{i}g_{i}}{\sum_{i=1}^{N}p_{i}^{2}+\sum_{i=1}^{N}g_{i}^{2}}, \tag{6}$$

where $P$ represents the predicted value, which is the probability result obtained by the Softmax operation, $P \in [None, 64, 128, 128, 4]$; $G$ stands for the mask, which is the ground truth after the one-hot encoding, $G \in [None, 64, 128, 128, 4]$. And N represents all voxels, $p_i \in P$ and $g_i \in G$. Here "None" represents the "batchsize" in the Tensorflow framework. The formula of "Categorical Dice" loss function is shown in (7), and the value of $\omega$ is [0.1, 1.0, 1.0, 1.0].

$$\text{Loss} = -2\frac{\sum_{l=1}^{4}\omega_{l}\sum_{i=1}^{N}p_{ln}g_{ln}}{\sum_{l=1}^{4}\omega_{l}\sum_{i=1}^{N}p_{ln}+g_{ln}}. \tag{7}$$

The partial differential operation can be performed on the formula (6) to obtain the calculated gradient relative to the predicted j-th voxel.

$$\frac{\partial D}{\partial p_{j}} = 2\left[\frac{g_{j}(\sum_{i}^{N}p_{i}^{2}+\sum_{i}^{N}g_{i}^{2})-2p_{j}(\sum_{i}^{N}p_{i}g_{i})}{(\sum_{i}^{N}p_{i}^{2}+\sum_{i}^{N}g_{i}^{2})^{2}}\right]. \tag{8}$$

## 4. Dataset and Experiments

*4.1. Dataset*

In this article, we use the 2020 BraTS challenge dataset to train and test our model. The dataset contains two types of brain tumor, i.e., high-grade glioblastoma (HGG) and low-grade glioma (LGG), each of which contains four modal images: fluid attenuation inversion recovery (FLAIR), T1 weighting (T1), T1-weighted contrast-enhanced (T1-CE), and T2 weighting (T2). Images of each modality can capture different characteristics of brain tumors. The mask of the brain tumor includes the necrosis area, edema area, enhancement area. Our task is to segment the sub-regions formed by the nesting of the three tags to enhance tumor (ET), whole tumor (WT), and tumor core (TC). We used the training set with 369 examples and 125 validation sets without labels provided by the BraTS Challenge, and the final accuracy was tested online on the official website.

*4.2. Evaluation metrics*

We use Dice coefficient, sensitivity, specificity, and Hausdorff95 distance to measure the performance of our model. Dice coefficient is calculated as:

$$\text{Dice} = \frac{2\text{TP}}{\text{FN} + \text{FP} + 2\text{TP}}, \tag{9}$$

where TP, FP, and FN are the number of true positive, false positive, and false negative respectively. Sensitivity can be used to evaluate the number of a true positive and false negative, it is used to measure the sensitivity of the model to segmented regions, defined as:

$$\text{Sensitivity} = \frac{\text{TP}}{\text{TP}+\text{FN}}. \tag{10}$$

Specificity can be used to evaluate the number of true negative and false positive, it is used to measure the model's ability to predict the background area, defined as:

$$\text{Specificity} = \frac{\text{TN}}{\text{TN}+\text{FP}}, \tag{11}$$

where TN is the number of true negatives. The Hausdorff95 distance measures the distance between the surface of the real area and the predicted area, which is more sensitive to the segmented boundary, defined as:

$$\text{Haus95}(T, P) = \max\{ \sup_{t \in T, p \in P} \inf d(t, p),\ \sup_{p \in P, t \in T} \inf d(t, p) \}, \tag{12}$$

where sup denotes the supremum and inf denotes the infimum, $t$ and $p$ donate the points on the surface $T$ of the ground-truth area and the surface $P$ of the predicted area, d $(\cdot, \cdot)$ is a function of the distance between the point *t* and *p*.

*4.3. Experimental detail*

In the field of deep learning, model training often involves the setting of hyperparameters, which can have a huge impact on the final training results of the model. However, how to set the value of the hyperparameter is often an empirical choice. When we train the ME-Net model, the initial learning rate is set to 0.0001, the dropout is set to 0.5, the number of training steps is about 450,000, and the data set is traversed 10 times. When training 200,000 steps, we reduce the learning rate to 0.00003, and to 0.00001 at 400,000 steps. After each traversal of the data set, we randomly shuffle the order of the data set to enhance the robustness of training.

The experimental environment is on Tensorflow with the runtime platform processor of Intel (R) Xeon (R) Silver 4210 CPU @ 2.20GHz 2.20GHz (2 processor) 128GB RAM, NVIDIA TITAN RTX, 64-bit Windows 10. The development software platform is PyCharm with Python 3.6.

## 5. Results and Discussion

*5.1. Results*

All 369 samples were used in the training process, and 125 samples were used for independent testing. The mask of the brain tumor includes the necrosis area, edema area, enhancement area, and background area. These tags are combined into three nested sub-regions: the enhancing tumor (ET), the whole tumor (WT), and the tumor core (TC). We use the four indicators of the Dice coefficient, sensitivity, specificity, and Hausdorff95 distance to measure the performance of the model. The average results of the indicators obtained on the training set and the validation set are shown in Table 1. The experimental results show that the model has good robustness. We observe that the model has a better segmentation effect on the WT region of brain tumors. The Dice of the WT area training set is 0.894, and the Dice of the validation set is 0.883, which is significantly better than the results of other sub-regions. The reason for this situation may be that the WT area has the largest range and is easier to segment. In addition, neurosurgeons, neuroradiologists, and radiologists have reported a high degree of uncertainty in the boundaries between ET and TC area labels (S. Bakas and H. Akbari et al., 2017). This is also the reason that the segmentation accuracy of ET and TC regions is lower than that of WT regions.

Table 1: The results of indicators of different partitions on the validation set.

|            | Dice  |       |       | Sensitivity |       |       | Specificity |       |       | Hausdorff95 |     |      |
|------------|-------|-------|-------|-------------|-------|-------|-------------|-------|-------|-------------|-----|------|
|            | ET    | WT    | TC    | ET          | WT    | TC    | ET          | WT    | TC    | ET          | WT  | TC   |
| Train      | 0.734 | 0.894 | 0.825 | 0.781       | 0.912 | 0.836 | 0.999       | 0.999 | 0.999 | 29.7        | 6.3 | 9.9  |
| Validation | 0.702 | 0.883 | 0.739 | 0.724       | 0.905 | 0.742 | 0.999       | 0.999 | 0.999 | 38.6        | 7.0 | 30.2 |

Fig. 5 and Fig. 6 respectively show the composite plot of the scatter plot and box plot of the evaluation index results of the training set and the validation set. It can be seen from the box plot that the fluctuation of the result is very small and only a few outliers appear. In the box plot, the horizontal line in the middle of the box represents the median of a set of data. We observe that the median is higher than the average, and the data results are slightly left-skewed, indicating that the segmentation results are concentrated in the higher area. It can also be observed in the scatter plot. In all the results, the specificity value has the smallest fluctuation range. This shows that the model has a strong ability to predict the background area. At the same time, the sensitivity has also reached a high range, sensitivity represents the model's ability to predict the segmented area. The smaller the difference between sensitivity and specificity, the better the model's performance in solving class imbalance. It can be seen from the figure below that the sensitivity fluctuation range is not large, especially the sensitivity and specificity of the WT region are the same, indicating that the model has the same ability to predict the foreground and the background, and can better solve the problem of the imbalance of foreground and background voxels. After verification, the overall performance of the model has reached a very high level.

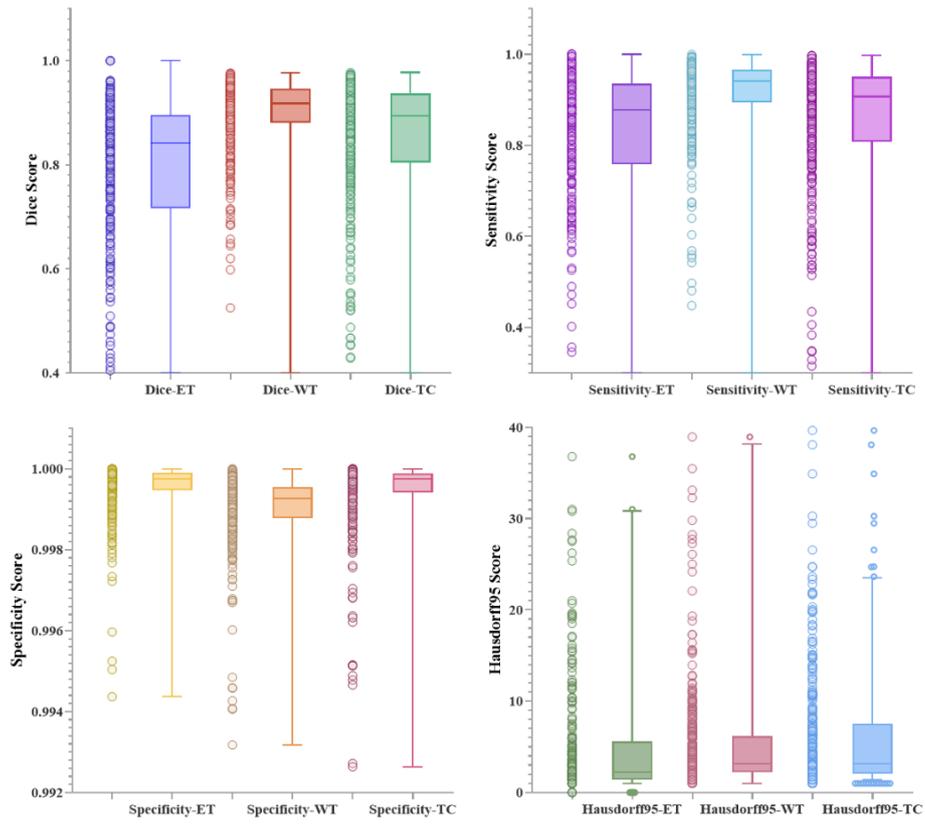

**Fig. 5.** Combination of scatters plot and box plot of the training set.

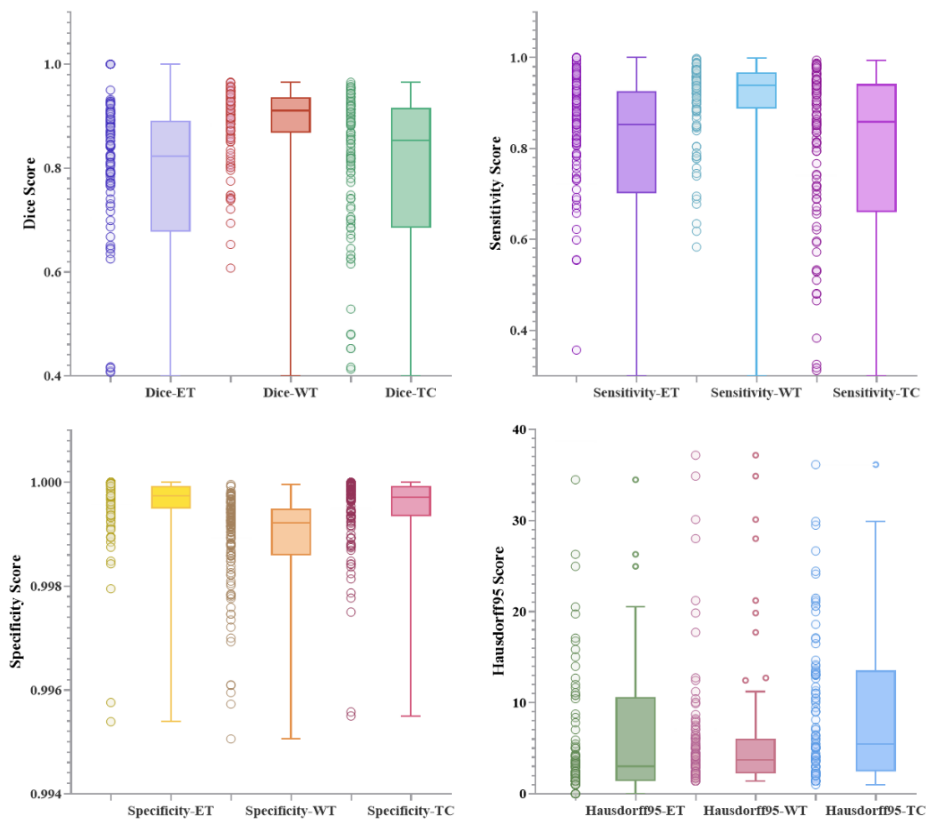

**Fig. 6.** Combination of scatters plot and box plot of the validation set.

We randomly selected several slices from the training set, and compared their actual situation with the predicted situation, as shown in Fig.7 (a). At the same time, we also selected two slices in Fig.7 (a) to show the monomodal results, as shown in Fig.7 (b). The red area represents the tumor core, the combined areas of yellow and red represent the enhancing tumor core, and the entire visible area, that is, the combined areas of green, yellow and red represent the whole tumor. From these segmentation examples, we noticed that the model has a good performance on the segmentation of brain tumor images, but the prediction of the tumor core is slightly biased. The reason may be that the area of the tumor core is too small, and the features are not obvious.

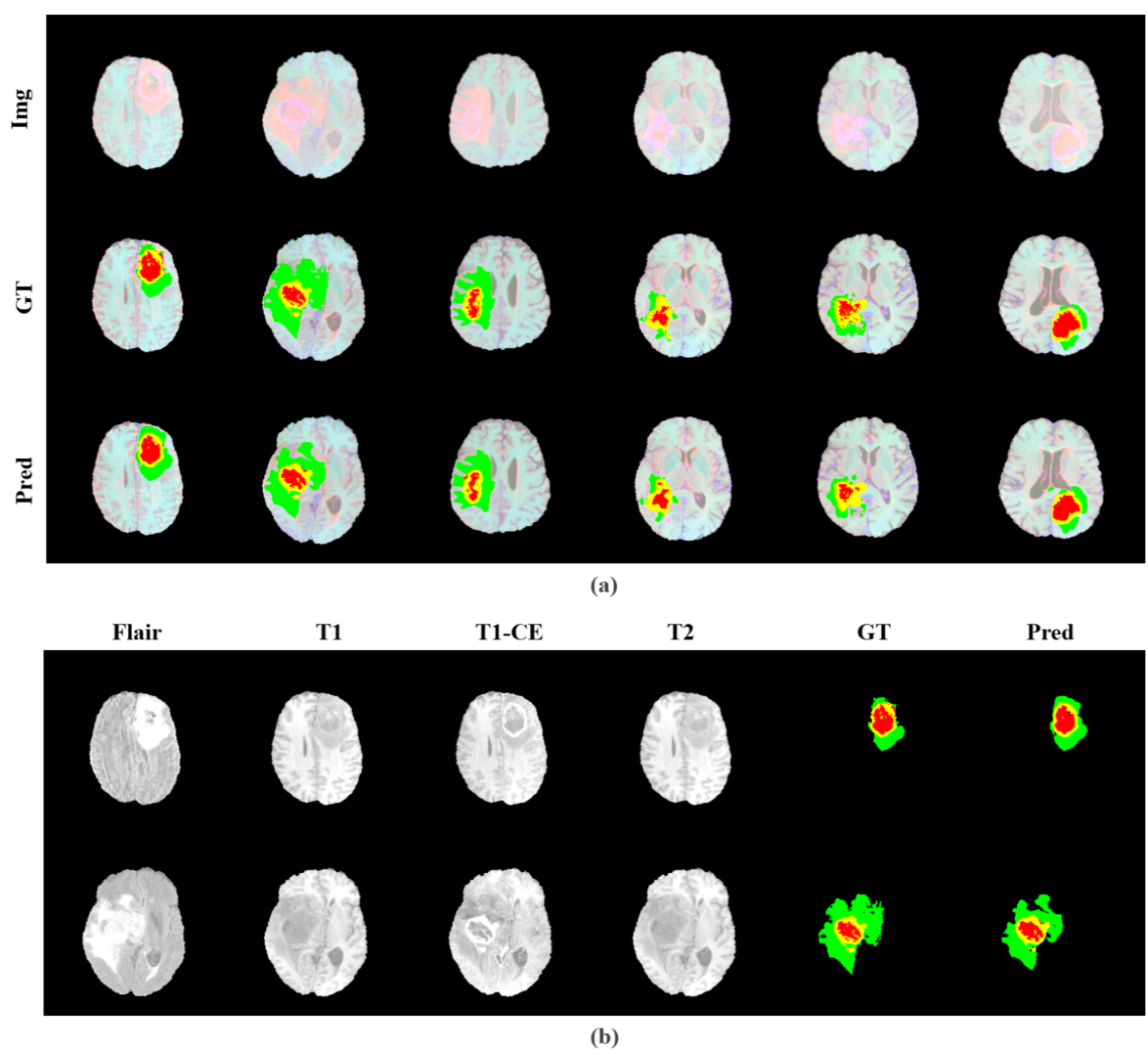

**Fig. 7.** Example of segmentation in the training set. In the figure, Img represents the original image, GT represents the label manually segmented by the expert, and Pred represents the result predicted by our model.

After the training is completed, we randomly select several segmentation slices and monomodal slices in the validation set for display, as shown in Fig. 8. And the Dice score of the ET area is marked on Fig.8 (b). From these slices, we can observe that the model has good segmentation results for brain tumors of different positions, sizes, and shapes. For MRI images of different intensities, it also has good segmentation results. It shows that our model can accurately separate lesion sub-regions, and can prove that this model has potential in segmenting brain tumor images.

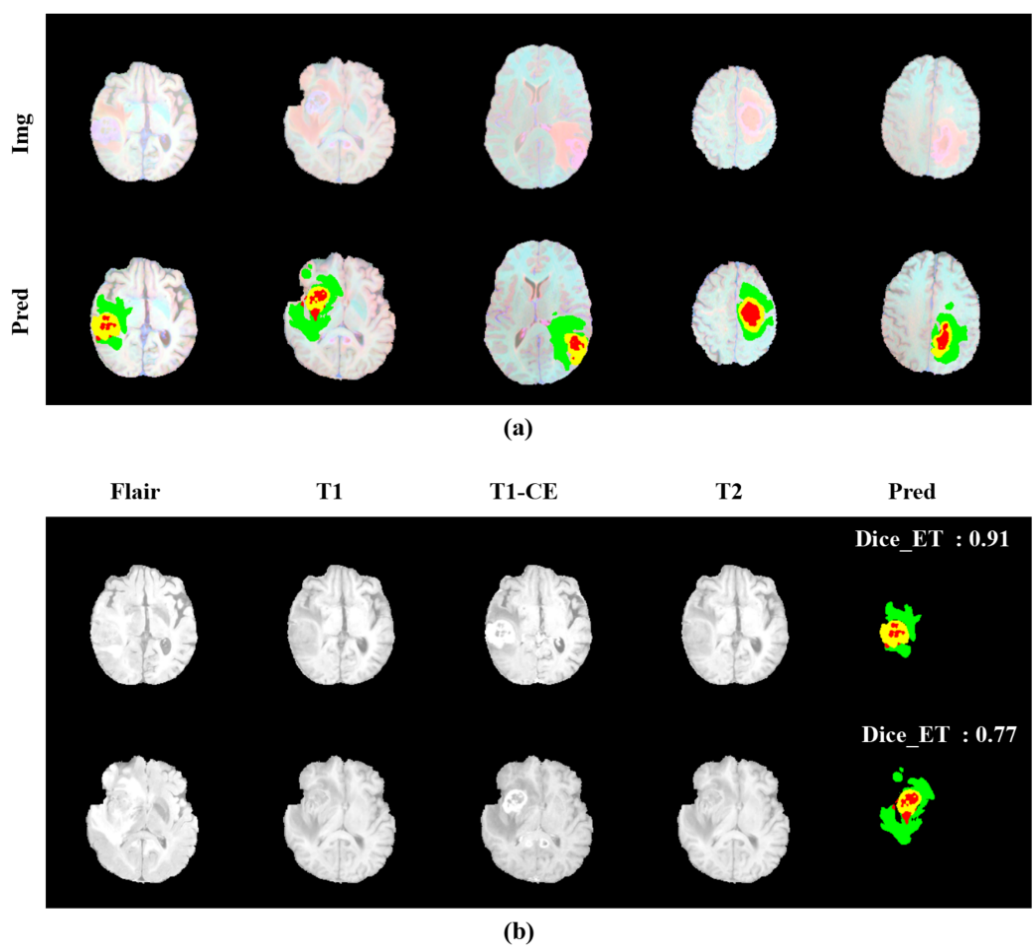

**Fig. 8.** Example of segmentation in the validation set. In the figure, Img represents the original image, Pred represents the result predicted by our model.

## 5.2. Comparison

In this article, we propose a new model of multi-encoders to segment 3D MRI brain tumor images and obtain better segmentation results on the BraTS 2020 dataset. In order to further demonstrate the effectiveness of this method, we compare the experimental results with other advanced methods and also compare with the results of the excellent team participating in the BraTS 2020.

Pereira et al. proposed an automatic segmentation method based on convolutional neural networks, using a small 3×3 kernel to suppress overfitting and allowing the use of deeper architectures, winning the second place in the BraTS 2015 challenge (Pereira et al., 2016). Chen et al. developed a 3D CNN-based model to automatically segment gliomas and extract context information at multiple scales. In addition, they used densely connected convolutional blocks to further improve performance (Chen et al., 2018). Aiming at the complexity of the model cascade strategy, Zhou et al. proposed a lightweight deep model OM-Net, which can better solve the problem of class imbalance, and designed a cross-task guided attention (CGA) module, in BraTS 2018 won the third place (Zhou et al., 2019). Zhao et al. integrated full convolutional networks and conditional random fields under a unified framework, and developed a new segmentation framework. First use image patches to train FCNNs (Zhao et al., 2017). Then use image slices with fixed FCNNs parameters to train CRFs as recurrent neural networks, and finally fine-tune FCNNs and CRF-RNNs. They evaluated the method at BraTS 2016.

The results compared with the above methods are shown in Table 2. Observing the results of the comparison shows that we have obtained the best results in the accuracy of the WT area. Since the method mentioned above is different from the data set used in this article, we compare it with other teams participating in the BraTS 2020 Challenge. We obtained the index results of other teams on the official website of the challenge, and the comparison results are shown in Table 3, from which we can observe comparable results compared to other state-of-the-art algorithms. The bold text in the table represents the best performance.

**Table 2:** Comparison of Dice results between the proposed method and different methods.

| Method | Dice_ET | Dice_WT | Dice_TC |
| --- | --- | --- | --- |
| **Proposed** | 0.70 | **0.88** | 0.74 |
| Pereira et al. | 0.65 | 0.78 | 0.75 |
| Chen et al. | **0.81** | 0.72 | **0.83** |

| | | 0.65 | 0.87 | 0.75 |
|---|---|---|---|---|
| | Zhou et al. | | | |
| | Zhao et al. | 0.62 | 0.84 | 0.73 |

**Table 3:** Comparison of teams participating in the BraTS 2020 challenge.

| | Team | Dice | | | Sensitivity | | | Specificity | | | Hausdorff95 | | |
|---|---|---|---|---|---|---|---|---|---|---|---|---|---|
| | | ET | WT | TC | ET | WT | TC | ET | WT | TC | ET | WT | TC |
| Train | **Proposed** | 0.74 | **0.89** | 0.83 | **0.78** | **0.91** | **0.84** | **0.99** | **0.99** | **0.99** | 29.7 | **6.3** | 9.9 |
| | Iris | 0.76 | 0.89 | 0.81 | 0.78 | 0.89 | 0.84 | 0.99 | 0.99 | 0.99 | 32.5 | 15.3 | 13.4 |
| | MARS | **0.78** | 0.88 | 0.83 | 0.78 | 0.84 | 0.80 | 0.99 | 0.99 | 0.99 | 29.3 | 6.23 | 7.0 |
| | CBICA | 0.54 | 0.78 | 0.57 | 0.64 | 0.82 | 0.53 | 0.99 | 0.99 | 0.99 | **20.0** | 46.3 | 39.6 |
| | agussa | 0.67 | 0.87 | 0.79 | 0.69 | 0.87 | 0.82 | 0.99 | 0.99 | 0.99 | 39.2 | 15.7 | 17.1 |
| | mpstanford | 0.60 | 0.78 | 0.72 | 0.56 | 0.80 | 0.75 | 0.99 | 0.99 | 0.99 | 36.0 | 17.7 | 17.2 |
| | unet3d | 0.69 | 0.81 | 0.75 | 0.77 | 0.93 | 0.83 | 0.99 | 0.99 | 0.99 | 37.7 | 19.6 | 18.4 |
| | LMB | 0.77 | 0.85 | **0.84** | 0.77 | 0.79 | 0.82 | 0.99 | 0.99 | 0.99 | 29.0 | 10.5 | **6.9** |
| | ovgu_seg | 0.58 | 0.79 | 0.65 | 0.68 | 0.79 | 0.69 | 0.99 | 0.99 | 0.99 | 47.3 | 20.0 | 23.3 |
| Validation | **Proposed** | 0.70 | **0.88** | 0.74 | 0.72 | **0.91** | 0.74 | **0.99** | **0.99** | **0.99** | 38.6 | **6.95** | 30.18 |
| | Iris | 0.71 | 0.88 | **0.77** | 0.72 | 0.90 | 0.75 | 0.99 | 0.99 | 0.99 | 35.2 | 15.5 | 20.6 |
| | MARS | **0.76** | 0.87 | 0.75 | **0.76** | 0.85 | 0.71 | 0.99 | 0.99 | 0.99 | 27.7 | 7.04 | **10.9** |
| | CBICA | 0.63 | 0.82 | 0.67 | 0.76 | 0.78 | 0.75 | 0.99 | 0.99 | 0.99 | **9.6** | 10.7 | 28.2 |
| | agussa | 0.59 | 0.83 | 0.69 | 0.60 | 0.87 | 0.71 | 0.99 | 0.99 | 0.99 | 56.6 | 23.2 | 30.0 |
| | mpstanford | 0.50 | 0.72 | 0.62 | 0.50 | 0.81 | 0.69 | 0.99 | 0.99 | 0.99 | 61.9 | 26.0 | 28.0 |
| | unet3d | 0.70 | 0.84 | 0.72 | 0.71 | 0.87 | **0.79** | 0.99 | 0.99 | 0.99 | 37.4 | 12.3 | 13.1 |
| | LMB | 0.72 | 0.82 | 0.76 | 0.70 | 0.77 | 0.72 | 0.99 | 0.99 | 0.99 | 37.4 | 12.3 | 13.1 |
| | ovgu_seg | 0.60 | 0.79 | 0.68 | 0.66 | 0.78 | 0.67 | 0.99 | 0.99 | 0.99 | 54.1 | 12.1 | 19.1 |

## 5.3. Discussion

### 5.3.1 Application areas

The segmentation method proposed in this paper is not affected by the size and location of the tumor, nor is it affected by the intensity of MRI. It can segment a variety of tumors and automatically identify healthy tissues, which is of great significance to the planning and arrangement of tumor surgery, radiotherapy, and other fields. The size of the tumor can also be displayed during segmentation, so the growth stage of the tumor can be determined to distinguish the treatment stage.

*5.3.2 Future work*

By analyzing the segmentation results of the model and comparing it with the top methods, we found that this method has some limitations. In this section, we list these limitations and possible solutions.

The segmentation effect of the ET region of the tumor is not very good. The reason may be that the characteristics of the ET region are blurred, and the features are not obvious. Inspired by Zhou et al., we will add some attention modules to the backbone network, such as the SE module, which consists of a global average pooling and two fully connected processes, through modeling network volumes (Hu et al., 2019). The interdependence between channels of product characteristics is used to selectively emphasize information characteristics and suppress less useful characteristics. Another example is the non-local module, which weighs the features of all locations, and this module can capture long-term dependencies (Wang et al., 2018). These modules can be transplanted to many computer vision architectures to recalibrate the features on the channel to perfect feature extraction.

Limited accuracy is a long-standing problem in the field of multi-class segmentation. Inspired by Chen et al., we expect to design a similar region clipping method, first find the part of interest, and then segment it. The above improvement methods may improve the performance of the model, but also increase the complexity of the algorithm. How to balance accuracy and complexity needs to be discussed in future experiments. In addition, according to the survey, different algorithms work best in different segmented sub-regions, but none of them is at the top of all sub-regions at the same time (B. H. Menze et al., 2015). How to integrate the advantages of various algorithms to achieve relatively high accuracy in each sub-region is also a direction for future brain tumor research.

Finally, one of the shortcomings we still have is that we haven't tried some better data enhancement methods. Good data enhancement can effectively prevent overfitting and enhance the generalization ability of the model. Andriy conducted experiments on the 2018 BraTS dataset and found some data enhancement methods that can improve accuracy, namely by applying random intensity offset and scaling and randomly flipping pictures on the input channel [(Andriy Myronenko, 2019)](#). In future work, we should combine previous studies to explore the impact of data enhancement on accuracy.

## 6. Conclusion

In this article, we proposed the ME-Net model, and performed experiments on the BraTS 2020 dataset, and obtained promising results. We designed four encoder structures for the four modal images of brain tumor MRI, and each modal corresponds to an encoder, which greatly improves the feature extraction ability in the downsampling process and improves the accuracy. At the same time, we use a large number of skip-connections to combine the feature maps of different modalities extracted by the four encoders. The final feature map is the feature module of the four encoders, and then the combined feature map is input to the decoder. We also introduced a new loss function, i.e., Categorical Dice, and set different weights for different masks at the same time. The weights for the three tumor regions of interest are set to 1, and the weight for the background of the larger area is set to 0. This method can well solve the problem of unbalanced foreground and background voxels that often occur in segmentation tasks. We evaluated our approach using an online verification tool on the BraTS Challenge website. The average Dice indices obtained on the validation dataset are promising and they were 0.70249, 0.88267 and 0.73864 for the enhanced tumor, the entire tumor and the tumor core, respectively.

**Conflict of interest**

The authors declare no conflict of interest.

**Acknowledgment**

This work is funded in part by the National Natural Science Foundation of China (Grant No.

62072413), and also supported in part by the AI for Health Imaging Award 'CHAIMELEON: Accelerating the Lab to Market Transition of AI Tools for Cancer Management' [H2020-SC1-FA-DTS-2019-1 952172].

**References**

S. Bakas, M. Reyes, A. Jakab, S. Bauer, M. Rempfler, A. Crimi, et al, 2018. Identifying the Best Machine Learning Algorithms for Brain Tumor Segmentation, Progression Assessment, and Overall Survival Prediction in the BRATS Challenge, arXiv preprint arXiv:1811.02629.

Sérgio Pereira, Adriano Pinto, Victor Alves, and Carlos A. Silva, 2016. Brain Tumor Segmentation Using Convolutional Neural Networks in MRI Images. IEEE Transactions on Medical Imaging, pp. 1240-1251.

Andriy Myronenko, 2019. 3D MRI Brain Tumor Segmentation Using Autoencoder Regularization. Lecture Notes in Computer Science, vol 11384.

Chenhong Zhou, Changxing Ding, Xinchao Wang, Zhentai Lu, and Dacheng Tao. (2019) One-Pass Multi-Task Networks With Cross-Task Guided Attention for Brain Tumor Segmentation. Computer Vision and Pattern Recognition.

B. H. Menze, A. Jakab, S. Bauer, J. Kalpathy-Cramer, K. Farahani, J. Kirby, et al. (2015) The Multimodal Brain Tumor Image Segmentation Benchmark (BRATS), IEEE Transactions on Medical Imaging, DOI: 10.1109/TMI.2014.2377694.

Feng Xue, Nicholas Tustison, and Craig Meyer. (2018) Brain Tumor Segmentation using an Ensemble of 3D U-Nets and Overall Survival Prediction using Radiomic Features. Lecture Notes in Computer Science, pp. 279-288.

Lele Chen, Yue Wu, Adora M.Dsouza, Anas Z.Abidin, Axel Wismuller, and Chenliang Xu. (2018) MRI Tumor Segmentation with Densely Connected 3D CNN. Image and Video Processing.

L.G. Nyul, J.K. Udupa and Xuan Zhang. (2000) New variants of a method of MRI scale standardization. IEEE Transactions on Medical Imaging, vol 19, pp. 143-150.

Jonathan Long, Evan Shelhamer, and Trevor Darrell. (2014) Fully Convolutional Networks for Semantic Segmentation. IEEE Transactions on Pattern Analysis & Machine Intelligence.

Olaf Ronneberger, Philipp Fischer, and Thomas Brox. (2015) U-Net: Convolutional Networks for Biomedical Image Segmentation. Computer Vision and Pattern Recognition.

Fausto Milletari, Nassir Navab, and Seyed-Ahmad Ahmadi. (2016) V-Net: Fully Convolutional Neural Networks for Volumetric Medical Image Segmentation. Computer Vision and Pattern Recognition.

Dmitry Lachinov, Evgeny Vasiliev, and Vadim Turlapov. (2018) Glioma Segmentation with Cascaded Unet. Computer Vision and Pattern Recognition.

Richard McKinley, Raphael Meier, and Roland Wiest. (2018) Ensembles of Densely-

Connected CNNs with Label-Uncertainty for Brain Tumor Segmentation. International Conference on Medical Image Computing and Computer Assisted Intervention (MICCAI 2018).

Chenhong Zhou, Shengcong Chen, Changxing Ding, and Dacheng Tao. (2019) Learning Contextual and Attentive Information for Brain Tumor Segmentation. Lecture Notes in Computer Science, vol 11384.

S. Bakas, H. Akbari, A. Sotiras, M. Bilello, M. Rozycki, J.S. Kirby, et al. (2017) Advancing The Cancer Genome Atlas glioma MRI collections with expert segmentation labels and radiomic features. Nature Scientific Data, 4:170117 DOI: 10.1038/sdata.2017.117.

Joseph K Leader, Bin Zheng, Robert M Rogers, Frank C Sciurba, Andrew Perez, Brian E Chapman, et al. (2003) Automated lung segmentation in X-ray computed tomography: development and evaluation of a heuristic threshold-based scheme1. Academic Radiology.

H Tang, E.X Wu, Q.Y Ma, D Gallagher, G.M Perera, T Zhuang. (2000) MRI brain image segmentation by multi-resolution edge detection and region selection. Computerized medical imaging and graphics.

Sulaiman S.N, Mat Isa N.A. (2010) Adaptive fuzzy-K-means clustering algorithm for image segmentation. IEEE Transactions on consumer electronics.

Lihong Juang, Mingni Wu. (2010) MRI brain lesion image detection based on color-converted K-means clustering segmentation. Measurement.

Wu X. (2015) An Iterative Convolutional Neural Network Algorithm Improves Electron Microscopy Image Segmentation.

Jie Hu, Li Shen, Samuel, and Albanie. (2019). Squeeze-and-excitation networks. IEEE transactions on pattern analysis and machine intelligence.

Xiaolong Wang, Ross Girshick, Abhinav Gupta, Kaiming He. (2018). Non-local Neural Networks.

Saddam Hussain, Syed Muhammad Anwar, and Muhammad Majid. (2017). Segmentation of Glioma Tumors in Brain Using Deep Convolutional Neural Network. Neurocomputing.

Graves A, Mohamed A. R., and Hinton G. (2003). Speech recognition with deep recurrent neural networks.

Stollenga M. F. , Byeon, W. , Liwicki, M. , and Schmidhuber J. (2015). Parallel multi-dimensional LSTM, with application to fast biomedical volumetric image segmentation. Computer ence.

Xiaomei Zhao, Yihong Wu, Guidong Song, Zhenye Li, Yazhou Zhang, and Yong Fan. (2017). A deep learning model integrating fcnns and crfs for brain tumor segmentation. Medical

Image Analysis, 43, 98-111.

Hussain S, Anwar S M, Majid M. (2017) Segmentation of Glioma Tumors in Brain Using Deep Convolutional Neural Network. Neurocomputing.

Yu Wang, Changsheng Li, Ting Zhu, and Chongchong Yu et al. (2019). A Deep Learning Algorithm for Fully Automatic Brain Tumor Segmentation. IEEE International Joint Conference on Neural Networks.

Marcel Prastawa, Elizabeth Bullitt, Sean Ho, and Guido Gerig. (2004). A brain tumor segmentation framework based on outlier detection. Medical Image Analysis.

Comelli Albert, Navdeep Dahiya, Alessandro Stefano, Viviana Benfante, Giovanni Gentile, Valentina Agnese, Giuseppe M. Raffa et al. (2020). Deep learning approach for the segmentation of aneurysmal ascending aorta." Biomedical Engineering Letters, 1-10.

Comelli Albert, Claudia Coronnello, Navdeep Dahiya, Viviana Benfante, Stefano Palmucci, Antonio Basile, Carlo Vancheri, Giorgio Russo, Anthony Yezzi, and Alessandro Stefano. (2020). Lung Segmentation on High-Resolution Computerized Tomography Images Using Deep Learning: A Preliminary Step for Radiomics Studies. Journal of Imaging 6, no. 11, 125.

Stefano Alessandro, Mauro Gioè, Giorgio Russo, Stefano Palmucci, Sebastiano Emanuele Torrisi, Samuel Bignardi, Antonio Basile et al. (2020). Performance of Radiomics Features in the Quantification of Idiopathic Pulmonary Fibrosis from HRCT. Diagnostics 10, no. 5, 306.

Sudre C.H., Li W., Vercauteren T., Ourselin S., Jorge Cardoso M. (2017). Generalised Dice Overlap as a Deep Learning Loss Function for Highly Unbalanced Segmentations. Lecture Notes in Computer Science, vol 10553. Springer, Cham.